\documentstyle[preprint,aps,graphicx]{revtex}

\begin{document}

\title{History and new ideas for exotic particles}

\author{Harry J. Lipkin}

\address{Department of Particle Physics,
Weizmann Institute of Science, \\ 
Rehovot, Israel \\ 
E-mail: harry.lipkin@weizmann.ac.il\\
and\\
School of Physics and Astronomy \\
Raymond and Beverly Sackler Faculty of Exact Sciences \\
Tel Aviv University, Tel Aviv, Israel\\
and\\
High Energy Physics Division, Argonne National Laboratory\\
Argonne, IL 60439-4815, USA}

\maketitle
\begin{abstract}

Basic 1966 physics of Sakharov,
Zeldovich and Nambu updated by QCD with constituent-quark quasiparticles 
having effective masses fits all  masses and
magnetic moments of ground state meson and baryons having no more than one
strange or heavy quark Flavor antisymmetry  
explains absence of
low-lying exotics and suggests diquark-triquark model and two-state model 
for $\Theta^+$ pentaquark. Variational approach gives mass bounds for other
pentaquarks.
\end{abstract}

\def\bra#1{\left\langle #1\right|}
\def\ket#1{\left| #1\right\rangle}

\section{Introduction - What can QED teach us about QCD?}

QCD is a Great Theory, but how do we  connect it with experiment or
find approximations   I recall Yoshio Yamaguchi's response 
in 1960 when asked whether there had been any thought at CERN
 about a possible breakdown of QED at small distances: ``No. .   Many
calculations. No thought."

What can we learn from QED; a Great Theory
that everyone knows how to connect with experiment?
We know how isolated free electrons behave and carry currents. 
But nobody could explain the fractional Hall effect until Robert Laughlin told
us  the Hall Current is not carried by single electrons 
but by quasiparticles related to electrons by a complicated
transformation.

Nobody has ever seen an isolated free quark. 
Experiments tell us that baryons are $qqq$ and mesons are $q \bar q$  
but these are not the current quarks whose fields appear in the QCD Lagrangian.
Are these quarks complicated quasiparticles related to current quarks
by a complicated transformation? Nobody knows.
Is Hadron Spectroscopy Waiting for Laughlin?
Does QCD need another Laughlin to tell us what constituent quarks are?

\section{ The 1966 basic physics of hadron spectroscopy}

\subsection{The QCD-updated Sakharov-Zeldovich mass formula}

A unified mass
formula for  both meson and baryon  ground state masses\cite{SakhZel} updated
by  DeRujula, Georgi and Glashow\cite{DGG}  (DGG) using QCD arguments  relating
hyperfine splittings to constituent quark effective masses\cite{Postcard} 
and baryon magnetic 
moments showed that
all are made of the same quarks\cite{SakhZel} and gave remarkable agreement with
experiment including three magnetic moment predictions with no free
parameters \cite{OlPenta,NewPenta}   

\begin{equation}
M = \sum_i m_i + \sum_{i>j}  
{{\vec{\sigma}_i\cdot\vec{\sigma}_j}\over{m_i\cdot
m_j}}\cdot v^{hyp}_{ij}
\label{sakhzel}
\end{equation}
\begin{eqnarray}
& \langle m_s-m_u \rangle_{Bar}&= 
M_\Lambda-M_N=177\,{\rm MeV}\nonumber \\[4pt]
&{}\langle m_s-m_u \rangle_{mes}& =
{{3(M_{K^{\scriptstyle *}}-M_\rho )
+M_K-M_\pi}\over 4} =180\,{\rm MeV} \\[4pt]
& \langle m_s-m_u \rangle_{Bar}&=
{{M_N+M_\Delta}\over 6}\cdot
\left({{M_{\Delta}-M_N}\over
{M_{\Sigma^{\scriptstyle *}}-M_\Sigma}} - 1 \right)
=190\,{\rm MeV}
\nonumber \\[4pt]
&{}\langle m_s-m_u \rangle_{mes}&=
 {{3 M_\rho + M_\pi}\over 8}
\cdot
\left({{M_\rho - M_\pi}\over{M_{K^*}-M_K}} - 1 \right)
= 178\,{\rm MeV}
\, ,
\label{ight:mass}
\end{eqnarray}
\begin{equation}
 \left({{m_s}\over{m_u}}\right)_{Bar} =
{{M_\Delta - M_N}\over{M_{\Sigma^*} - M_\Sigma}} = 1.53 = 
 \left({{m_s}\over{m_u}}\right)_{mes}= 
{{M_\rho - M_\pi}\over{M_{K^*}-M_K}}= 1.61
\end{equation}

\begin{eqnarray}
&&{} \mu_\Lambda=
-0.61 
\,{\rm n.m.}=
-{\mu_p\over 3}\cdot {{m_u}\over{m_s}} =
-{\mu_p\over 3} {{M_{\Sigma^*} - M_\Sigma} \over{M_\Delta - M_N}}
=-0.61 \,{\rm n.m.}
\nonumber \\[4pt]
&&{}
\mu_p+\mu_n= 0.88 \,{\rm n.m.}
={M_{\scriptstyle p}\over 3m_u}
={2M_{\scriptstyle p}\over M_N+M_\Delta}=0.865 \,{\rm n.m.}
\nonumber \\[4pt]
&&{}
-1.46 =
{\mu_p \over \mu_n} =
-{3 \over 2}\, ,
\label{mag:mom}
\end{eqnarray}

 The same value $\pm 3\%$ for $m_s-m_u$ is obtained from 
four independent calculations. 
The same value $\pm 2.5\%$ for ${{m_s}\over{m_u}}$ is obtained from  
meson and baryon masses.
The 
same approach for $m_b-m_c$  gives
\begin{eqnarray}
&& \langle m_b-m_c \rangle_{Bar}= M(\Lambda_b)-M(\Lambda_c) =3341 \,{\rm MeV}
\nonumber \\[4pt]
&& \langle m_b-m_c \rangle_{mes} =
{{3(M_{B^{\scriptstyle *}}-M_{D^{\scriptstyle *}})
+M_B-M_D}\over 4} =3339 \,{\rm MeV} 
\label{heavy:m}
\end{eqnarray}

\subsection{Two Hadron Spectrum  puzzles -Why $qqq$ and $q\bar q$ ? }

\begin{enumerate}

\item The Meson-Baryon Puzzle - The $qq$ and $\bar qq$ forces bind both mesons
and baryons differently. A vector interaction gives equal and opposite
forces; a scalar or tensor gives equal attractions for both.

\item Exotics Puzzle - No low-lying hadrons with exotic quantum numbers have 
been observed; e.g. no $\pi^+ \pi^+$ or $K^+ N$ bound states.
\end{enumerate} 

Nambu solved both puzzles\cite{Nambu} in 1966, related mesons and baryons and
eliminated exotics by introducing color and a two-body non-abelian-gauge
interaction with the color-factor of one-gluon exchange. 

A unified treatment of $qq$ and $\bar qq$ interactions binds both mesons and
baryons with the same forces.  
Only $qqq$ and $q\bar q$ are stable in any single-cluster model with color
space factorization. Any color singlet cluster that can break up into two color singlet 
clusters loses no color electric energy and gains kinetic energy.
The Nambu color factor does not imply dynamics of one-gluon exchange. 
Higher order diagrams can have same color factor

Looking beyond bag or single-cluster models for possible molecular bound
states  Lipkin(1972) lowered  the color-electric potential energy   in potential models by introducing color-space correlations; e,g,
$q\bar qq \bar q$ at  corners of a square, but not enough to compensate for the
kinetic energy\cite{LipkTriEx}
\subsection{Important systematics in the experimental spectrum}
A large spin-dependent interaction $\approx$ 300 MeV
but a very weak interaction $\approx$ 2 MeV binding normal hadrons.
\begin{equation}
 M(\Delta) - M(N) \approx 300 MeV \gg M(n) + M(p) - M(d)
 \approx 2 MeV 
\end{equation}
\subsection{Conclusions from basics - What we do know and don't}

We know the low-lying hadron spectrum is described by quasiparticles called quarks with
a linear effective mass term and  a hyperfine interaction with a one-gluon
exchange color factor.
Only color singlet and $3^*$ color factors
arise in
the $(\bar q q)$ and $(qqq)$ states which behave like neutral atoms
with a strong color electric field inside hadrons and none outside. 
No molecular bound states arise in the simplest cases.
A strong spin-dependent interaction is crucial to understanding the spectrum

We don't know what these quarks are and the low-lying hadron spectrum provides 
no direct experimental information on $(\bar q q)_8$ and $(qq)_6$  interactions
needed for multiquark exotic configurations.

\section{QCD Guide to the search for exotics}
\subsection{Words of Wisdom from Wigner and Bjorken}
Wigner said:  ``With a few free parameters I can fit an elephant. 
With a few more I can make him wiggle his trunk"

His response to questions about a particular theory he did not like was:

``I think that this theory is wrong. But  the old Bohr - Sommerfeld quantum
theory  also wrong.  It is hard to see how we could have reached the
right theory  without going through that stage'.

  In 1986 Bjorken noted how a $q \bar q$  created in $e^+ e^- $ annihilation
fragments into hadrons.  The quark can pick up an antiquark to make a meson. 
or a quark to make diquark. The diquark can pick up another quark to make a
baryon  but might pickup an antiquark to make a ``triquark" bound in a color
triplet state. Picking up two more quarks makes a pentaquark

BJ asked: ``Should such states be bound or live long enough to be observable as
hadron resonances? What does quark model say?

\subsection{What the quark model says about exotics}
 
To consider the possible mass difference between the $\Theta^+$ and a separated
KN system, first put a $K^+$ and a neutron close together and keep the $u \bar
s$ in the kaon and the $udd$ in the neutron coupled to color singlets. Nothing
happens because color singlet states behave like neutral atoms with 
negligible  new interactions. Next change   color-spin couplings while keeping
an overall color singlet and search for the minimm energy. Use a variational
approach with wave functions having the same spatial two-body density matrix
elements as those in the observed mesons and baryons.  Experimental hadron mass
differences are then used to determine all parameters and look for possible
bound states.

 This approach finds no possibility for a $K^+n$ bound state. But the same
method   shows that   this trial wave function for the $D_s^-p$ system gives a
lower  hyperfine potential energy for the anticharmed strange pentaquark ($\bar
c suud$)     over  the separated $D_s^-p$.  Whether this is enough to
compensate for the kinetic energy required to localize the state is unclear and
highly model dependent with too many unknown parameters as soon as  the
requirements on the two-body density matrix are relaxed . 

 This anticharmed strange pentaquark\cite{Pent97} and Jaffe's H
dibaryon\cite{Jaffe} became the subjects of experimental searches. Although
Fermilab E791 did not find convincing evidence\cite{E791Col} for the $\bar c
suud$ pentaquark, the possibility is still open that this stable bound
pentaquark exists and needs a better search.

      The existence of the $\Theta^+$ showed that wave
functions with the same two-body density matrix for all pairs did not work
and a two cluster model was  needed to separate the $uu$ and $dd$ pairs that
have a repulsive short-range hyperfine interaction. This led to the
diquark-triquark model\cite{OlPenta,NewPenta}..

\subsection{Crucial role of color-magnetic interaction}
\begin{enumerate}

\item QCD motivated models show same color-electric interaction  for large 
multiquark states  and separated hadrons and no binding. Only  short-range
color-magnetic interaction produces binding.

\item  Jaffe\cite{Jaffe} (1977) extended DGG with same color factor to
multiquark sector in a single cluster or bag model, defined $(\bar qq)_8$ and
$(qq)_6$ interactions, explained absence of lowlying exotics and 
suggested search for $H$ dibaryon $uuddss$. 

\item Jaffe's model extended to heavy quarks and
flavor-antisymmetry principle\cite{liptet} suggested exotic tetraquarks and
anticharmed strange pentaquark\cite{Pent97} $(\bar cuuds)$ (1987)

\end{enumerate}
\subsection{Flavor antisymmetry principle - No leading exotics}

The Pauli
principle requires flavor-symmetric quark pairs 
to be antisymmetric in color
and spin at short distances. Thus the short-range color-magnetic interaction is always
repulsive between flavor-symmetric pairs. 

\begin{enumerate}

\item Best candidates for multiquark binding have minimum number of
same-flavor pairs

\begin{enumerate}

\item Nucleon has only one same-flavor pair  

\item $\Delta^{++} (uuu)$ has three same-flavor pairs  

Costs 300 Mev relative to nucleon with only one.

\item Deuteron  separates six same-flavor pairs into two nucleons
 
 Only two same-flavor pairs  feel short range repulsion.  

\item $H (uuddss)$ has three same-flavor pairs. Optimum for light 
quark dibaryon

\item The $(uuds \bar c)$ pentaquark has only one same-flavor pair
\end{enumerate}

\item Pentaquark search. $(uuds \bar c)$ pentaquark has same
 binding  as H. 
\begin{enumerate} 
\item Quark model calculations
told experimenters to look for  $(uuds \bar c)$ pentaquark; not the 
$\Theta^+$. 

\item $\Theta^+$ $(uudd \bar s)$ has two same-flavor pairs pairs.
  Too many for a single baryon. 

\item Calculations motivating the 
$(uuds \bar c)$ pentaquark search 
found no reason to look for $(uudd \bar s)$
\end{enumerate}
\end{enumerate}

 Ashery's E791 search for $\bar c uuds$ found events\cite{E791Col}; not 
convincing enough. 

Better searches for this pentaquark are needed; e.g.
searches with good vertex detectors and particle ID\cite{Pent97}.

Any proton emitted from secondary vertex is interesting.
One gold-plated event  not a known baryon is enough; No
statistical analysis needed.

\section {The $\Theta^+$ was found! What can it be?} 

Following Wigner's guidance  to understand QCD and the pentaquark,
find a good wrong model that can teach us; stay away from 
free parameters 
 \subsection {The skyrmion model}
Experimental search motivated by another ``wrong model". 
Skyrmion\cite{DPP} has no simple connection with quarks 
except by another ``wrong model".  The $1/N_c$ expansion
invented\cite{LipHouch} pre-QCD to explain absence of free quarks.

-The 
binding Energy of $q \bar q$ pairs into mesons $ E_M \approx g^2 N_c$.

At large $N_c$ the cross section for meson-meson scattering breaking up a meson
into its constituent quarks is
\begin{equation}
\sigma [MM \rightarrow M + q + \bar q] \approx g^2  {{E_M}\over{N_c}}
\approx 0 
\end{equation}
But ${1\over {N_c}} ={1\over 3}$;  ${\pi\over {N_c}} \approx 1$  
This is NOT A SMALL PARAMETER!
\subsection {How to explain $\Theta^+$ with quarks - The two-state model}
 No bag or single cluster model with the same flavor-space correlation  for all
quarks can work. Keeping same-flavor pairs apart led 
to diquark-triquark model with  $(ud)$  diquark separated from  
remaining $(ud \bar s)$ triquark with triquark color-spin coupling minimizing
color-magnetic energy \cite{OlPenta,NewPenta}.

Noting two different color-spin couplings for triquark
with roughly equal color-magnetic energy leads naturally to a two-state 
model\cite{Karliner:2004qw}.

Let $\ket {\Theta_1}$ and $\ket {\Theta_2}$ denote an orthonormal basis  for
the two diquark-triquark states with different triquark color-spin couplings.

The mass matrix eigenstates can be defined with a mixing angle $\phi$ 
\begin{eqnarray}
 &\ket \Theta_S &\equiv 
\cos \phi \cdot \ket {\Theta_1} +
\sin \phi \cdot \ket {\Theta_2}
 \nonumber \\[4pt]
& \ket \Theta_L & \equiv 
\sin \phi \cdot \ket {\Theta_1} -
\cos \phi \cdot \ket {\Theta_2} 
\label{eq:twostate}
\end{eqnarray}

Loop diagram via the $KN$ intermediate state 
$\Theta_i \rightarrow KN \rightarrow \Theta_j$
gives the mass matrix and mass eigenstates
\begin{equation}
M_{ij} = M_o\cdot \bra{\Theta_i} T \ket {KN}
\bra{KN} T \ket {\Theta_j}
\label{eq:mass}
\end{equation}
\begin{eqnarray}
 &
\ket \Theta_S &= C[
\bra{KN} T \ket {\Theta_1} \cdot \ket {\Theta_1}
+
\bra{KN} T \ket {\Theta_2} \cdot \ket {\Theta_2}]
\nonumber \\[4pt]
&\ket \Theta_L &= C[
\bra{KN} T \ket {\Theta_2} \cdot \ket {\Theta_1}
-
\bra{KN} T \ket {\Theta_1} \cdot \ket {\Theta_2}
]
\label{eq:eigenstate}
\end{eqnarray}
where $C$ is a normalization factor  

\noindent Then  $ \bra{KN} T \ket
{\Theta_1} \cdot \bra{KN} T \ket {\Theta_2} - \bra{KN} T \ket {\Theta_2} \cdot
\bra{KN} T \ket {\Theta_1}  = 0 $

\noindent Thus
 $ \bra{KN} T \ket \Theta_L = 0$; 
the state $\Theta_L$ is decoupled from $KN$ and its decay into 
$KN$ is forbidden.

The state $\Theta_S$ with normal hadronic width can escape observation
against continuum background.

But there are no restrictions on couplings  to $K^*N$. Both 
$\ket
{\Theta_L}$ and $\ket {\Theta_S}$ are produced without suppression by
$K^*$ exchange. 

 Advantages of the two-state model
\begin{enumerate}

\item Explains narrow width and strong production

\item Arises naturally in a diquark-triquark model 

where two states have different color-spin couplings

\item Loop diagram mixing via $KN$ decouples one state from $KN$

\item Broad state decaying to $KN$ not seen

\item Narrow state coupled weakly to $KN$ produced via $K^*$ exchange 

\end{enumerate}

\subsection { A variational approach for the Pentaquark 
Multiplet}
Apply the QM Variational Principle to the exact (unknown)
hamiltonian $H$   and  unknown exact wave function $\ket{\Theta^+} $
with three simple assumptions\cite{varxistar}:
\begin{enumerate}
\item Assume $\Theta^+$ and $\Xi^{--}$ 
are pentaquarks $uudd\bar s$ and $ssdd \bar u$ 
\item Assume $\Theta^+$ and $\Xi^{--}$ are degenerate in $SU(3)_f$ limit.
\item Assume  SU(3) breaking changes only quark masses   
and leaves QCD color couplings unchanged in $H$.
\end {enumerate}
 \begin{equation}
\bra{\Theta^+}T^{\dag}_{u\leftrightarrow s} 
 H T_{u\leftrightarrow s} 
- H \ket{\Theta^+} \approx
 m_s - m_u + \bra{\Theta^+}\delta V^{hyp}_{\bar s \rightarrow \bar u}+
\delta V^{hyp}_{u \rightarrow s} \ket{\Theta^+} 
\label{eq:SU3br}
\end{equation}
where the $SU(3)_f$ transformation $T_{u\leftrightarrow s}$ 
interchanges  $u$ and $s$ flavors and 
$\delta V^{hyp}_{\bar s \rightarrow \bar u}$ and  
$\delta V^{hyp}_{u \rightarrow s}$ denote the change in the hyperfine
interaction under the transformations $\bar s \rightarrow \bar u$ and
$u \rightarrow s$ respectively.
Define a trial wave function 
\begin{equation} 
\ket{\Xi^{--}_{var}} \equiv T_{u\leftrightarrow s} 
\cdot \ket{\Theta^+}  
\end{equation}
The variational Principle gives an upper bound for $ M(\Xi^{--})$ 
 \begin{eqnarray}
&M(\Xi^{--})  \leq \bra{\Xi^{--}_{var}} H \ket{\Xi^{--}_{var}}& =
\bra{\Xi^{--}_{var}} H \ket{\Xi^{--}_{var}}
 + M(\Theta^+)  -
\bra{\Theta^+} H \ket{\Theta^+}
\nonumber \\[4pt]
& M(\Xi^{--}) - M(\Theta^+)& \leq 
\bra{\Theta^+}T^{\dag}_{u\leftrightarrow s}  H T_{u\leftrightarrow s} 
- H \ket{\Theta^+} 
\nonumber \\[4pt]
& M(\Xi^{--}) - M(\Theta^+)& \leq  
m_s - m_u + 
\bra{\Theta^+}\delta V^{hyp}_{\bar s \rightarrow \bar u} 
+\delta V^{hyp}_{u \rightarrow s}\ket{\Theta^+} \, ,  
\label{eq:var}
\end{eqnarray}
where we have substituted eq. (\ref{eq:SU3br}) for the $SU(3)_f$ breaking 
piece of $H$.
From quark model hadron spectroscopy and
simple assumptions about SU(3) breaking 
 \begin{eqnarray}
 &m_s - m_u & \leq M(\Lambda) - M(N) 
\nonumber \\[4pt]
&\bra{\Theta^+}\delta V^{hyp}_{\bar s \rightarrow \bar u}\ket{\Theta^+}
 &\leq 0 
 \nonumber \\[4pt]
&\bra{\Theta^+}\delta V^{hyp}_{u \rightarrow s} \ket{\Theta^+} &\leq
 2 \cdot\bra{ud_{S=0}} \delta V^{hyp}_{u \rightarrow s} \ket{ud_{S=0}}
\, ,
\label{eq:SU3br2}
\end{eqnarray}

 \begin{eqnarray}
 &M(\Sigma^-)-M(\Lambda)& = 
 \bra{ud_{S=0}} \delta V^{hyp}_{u \rightarrow s} \ket{ud_{S=0}} -
\bra{ud_{S=1}} \delta V^{hyp}_{u \rightarrow s} \ket{ud_{S=1}}
 \nonumber \\[4pt]
&& = \bra{ud_{S=0}} \delta V^{hyp}_{u \rightarrow s} \ket{ud_{S=0}}\cdot
\left( 1 - {{\bra{ud_{S=1}} \delta V^{hyp}_{u \rightarrow s} \ket{ud_{S=1}}}\over{ 
\bra{ud_{S=0}} \delta V^{hyp}_{u \rightarrow s} \ket{ud_{S=0}}}} \right )
 \nonumber \\[4pt]
 &M(\Sigma^{*-})-M(\Delta^o)& = 
 m_s - m_u + 2\bra{ud_{S=1}}\delta V^{hyp}_{u \rightarrow s}
  \ket{ud_{S=1}} 
\label{eq:hadspec}
\end{eqnarray}

 \begin{eqnarray}
 &
M(\Xi^{--}) - M(\Theta^+) &\leq 
m_s - m_u + 
 2 \cdot\bra{ud_{S=0}} \delta V^{hyp}_{u \rightarrow s} \ket{ud_{S=0}}
 \nonumber \\[4pt]
&M(\Xi^{--}) - M(\Theta^+) &\leq 
  M(\Sigma^{*-})-M(\Delta^o) +2[M(\Sigma^-)-M(\Lambda)] 
 \nonumber \\[4pt]
  &M(\Xi^{--}) - M(\Theta^+) &\leq  316 \rm{MeV} 
 \nonumber \\[4pt]
&[M(\Xi^{--}) - M(\Theta^+)]_{(experiment)}&=  330 \rm{MeV}
\label{eq:bound1}
\end{eqnarray}

\noindent With $\sigma_i \cdot \sigma_j$ interaction, ${{\bra{ud_{S=1}} \delta V^{hyp}_{u \rightarrow s} 
 \ket{ud_{S=1}}}\over{ 
\bra{ud_{S=0}} \delta V^{hyp}_{u \rightarrow s} \ket{ud_{S=0}}}} = -{1\over 3}$
\begin{eqnarray}
 &M(\Xi^{--}) - M(\Theta^+) &\leq 
M(\Lambda) - M(N) +{3\over 2}\cdot[M(\Sigma^-)-M(\Lambda)] 
 \nonumber \\[4pt]
 &M(\Xi^{--}) - M(\Theta^+) &\leq 299  \rm{MeV}  
 \nonumber \\[4pt]
&[M(\Xi^{--}) - M(\Theta^+)]_{(experiment)}&=  330 \rm{MeV} 
\label{eq:bound2}
\end{eqnarray}

Experiment violates both bounds!

 Is an experiment or one of our assumptions wrong?

\begin{enumerate} 
\item $\Theta^+$ and $\Xi^{--}$ 
not  pentaquarks $uudd\bar s$ and $ssdd \bar u$? 

\item $\Theta^+$ and $\Xi^{--}$ not degenerate in the $SU(3)_f$ limit?

\item Is our SU(3)-breaking model wrong? 
\end{enumerate} 

One possibility is  the two-state model.
The $\Theta^+$ and $\Xi^{--}$  are not in the same $SU(3)_f$ multiplet
if the 
two nearly degenerate diquark-triquark multiplets mix differently

\section {Heavy flavor pentaquaraks -
The $\Theta_c$ charmed pentaquark}
 
We now use the variataional approach to examine pentaquark states
obtained by replacing the $\bar s$ by $\bar c$ or other heavy aniquarks
in the exact  $\Theta^+$ wave function\cite{Karliner:2003si} and define
a trial wave function 

\begin{equation} 
\ket{\Theta_c^{var}} \equiv T_{s\leftrightarrow c} 
\cdot \ket{\Theta^+}  
\end{equation}

We have the same light quark system  and a different flavored antiquark.
There is the same color electric field and a mass change.  

The variational principle gives an upper bound for $ M(\Theta_c)$ 
\begin{equation}  
M(\Theta_c) \leq  M(\Theta^+)+ m_c - m_s + 
\langle V_{hyp}(\bar c) \rangle_{\Theta^+} -
\langle V_{hyp}(\bar s)\rangle_{\Theta^+}  
\end{equation}
Hyperfine interaction inversely proportional to quark mass product,
\begin{equation}
 \langle V_{hyp}(\bar c) \rangle =
{{m_s}\over{m_c}}\cdot\langle V_{hyp}(\bar s)\rangle
\end{equation}
\begin{equation} M(\Theta_c) - M(\Theta^+) \leq  (m_c - m_s)\cdot
\left(1+{{\mid\langle 
 V_{hyp}(\bar s) \rangle_{\Theta^+}\mid} \over{m_c}} \right) 
 \end{equation}

Now examine the difference between the mass and the decay threshold
\begin{eqnarray}
&\Delta E_{DN} (\Theta_c) &= M(\Theta_c) - M_N  - M_D  
\nonumber \\[4pt]   
&\Delta E_{KN} (\Theta^+) &=  M(\Theta^+) - M_N - M_K \approx 100 \rm{MeV} 
\nonumber \\[4pt]  
&\Delta E_{DN} (\Theta_c) - \Delta E_{KN} (\Theta^+) &= 
  M(\Theta_c) - M(\Theta^+) - M_D  + M_K 
\label{eq:threshold}
\end{eqnarray}
\begin{equation}
  \Delta E_{DN} (\Theta_c) \leq  
  0.7 \cdot\mid\langle V_{hyp}(\bar s) \rangle_{\Theta^+}\mid 
 -100~\rm{MeV} 
 \end{equation}
 Thus if  $\mid\langle 
 V_{hyp}(\bar s) \rangle_{\Theta^+}\mid \leq 140 ~\rm{MeV}$ the $\Theta_c$
is stable against strong decays. 

But the $K^*-K$ mass difference tells us that in the kaon 
\begin{equation}
 \mid\langle V_{hyp}(\bar s) \rangle_{K(u\bar s}\mid \approx 300 ~\rm{MeV}
 \end{equation} 
Is the hyperfine interaction of  $\bar s$ with four quarks in a $\Theta^+$
comparable to $V_{hyp}(\bar s)$ with one quark in a kaon?

This determines the stability of the $\Theta_c$.
Experiment will tell us about how QCD makes hadrons from quarks and gluons

\section{Experimental contradictions about the $\Theta^+$}  

Some experiments see the pentaquark
\cite{Nakano:2003qx}-\kern-0.5em others definitely do not\cite{cryptopen}.
 No theoretical model addresses  why
certain experiments see it and others do not. Comprehensive
review\cite{jenmalt} analyzes different models.

Further analysis is needed to check presence of specific production mechanisms
in experiments that see the $\Theta^+$ and their absence in those that
do not\cite{cryptopen}. One possibility is production and decay of a
cryptoexotic $N^*(2400)$ with hidden 
strangeness\cite{Landsberg:1999wn}
fitting naturally into 
$P$-wave $(ud)$ diquark-$ud \bar s$ triquark model for
the $\Theta^+$. The $N^*$ is a $(ds)$ diquark in the same flavor $SU(3)$ 
multiplet as the $(ud)$ diquark in the $\Theta^+$ in a $D$-wave with the $ud
\bar s$ triquark. Its dominant decay would produce the $\Theta^+$ in $K^-
\Theta^+$  via the diquark transition $ds \rightarrow ud + K^-$. Decays like
$\Lambda K$and $\Sigma K$ would be suppressed by the centrifugal barrier
forbidding a quark in the triquark from joining the diquark.

\section*{Acknowledgments}
The original work reported in this talk was in collaboration with Marek
Karliner.
This work was partially
supported by the U.S. Department
of Energy, Division of High Energy Physics, Contract W-31-109-ENG-38

\end{document}